\documentclass[preprint]{aastex}
\usepackage{lscape}

\newcommand\plotthree[3]{%
 \centering
 \leavevmode
 \includegraphics[width=.3\columnwidth]{#1}%
 \hfil
 \includegraphics[width=.3\columnwidth]{#2}%
 \hfil
 \includegraphics[width=.3\columnwidth]{#3}%
}%

\begin{document}

%% LaTeX will automatically break titles if they run longer than
%% one line. However, you may use \\ to force a line break if
%% you desire.

\title{Keck Diffraction-Limited Imaging of the Young \\ Quadruple
Star System HD 98800\altaffilmark{1}}

%% As in the title, you can use \\ to force line breaks.

\author{L. Prato\altaffilmark{2} and A. M. Ghez\altaffilmark{3}}
\affil{Department of Physics and Astronomy, UCLA, Los Angeles, CA 90095-1562}

\author{R. K. Pi\~na, C. M. Telesco, and R. S. Fisher}
\affil{Department of Astronomy, University of Florida,
Gainesville, FL 32611}

\author{P. Wizinowich, O. Lai, D. S. Acton, and P. Stomski}
\affil{W. M. Keck Observatory, 65-1120 Mamalahoa Hwy,
Kamuela, HI 96743}

%%\bigskip

%%\noindent
%%{To appear in the {\it Astrophysical Journal}}

%% Notice that each of these authors has alternate affiliations, which
%% are identified by the \altaffilmark after each name.  Specify alternate
%% affiliation information with \altaffiltext, with one command per each
%% affiliation.

\altaffiltext{1}{To appear in the {\it Astrophysical Journal}}
\altaffiltext{2}{lprato@astro.ucla.edu}
\altaffiltext{3}{Packard Fellow}

%% etc...

\begin{abstract}

This paper presents diffraction-limited 1$-$18$\mu$m images
of the young quadruple star system
HD 98800 obtained with the W. M. Keck 10-m telescopes
using speckle and adaptive optics imaging
at near-infrared wavelengths and
direct imaging at mid-infrared wavelengths.
The two components of the visual binary, A and B,
both themselves spectroscopic binaries, were separable
at all wavelengths, allowing us to determine their stellar and
circumstellar properties.  Combining these observations with
spectroscopic data from the literature, we derive an age of
$\sim$10$^7$ years, masses of 0.93 and 0.64 M$_{\sun}$ and an
inclination angle of 58$^{\circ}$ for the spectroscopic components of
HD 98800 B, and an age of $\sim$10$^7$ years
and a mass of 1.1 M$_{\sun}$ for HD 98800 Aa.  Our data confirm that
the large mid-infrared excess is entirely associated with
HD 98800 B.  This excess exhibits a black body
temperature of 150 K and
a strong 10$\mu$m silicate emission feature.  The theoretical equilibrium
radius of large, perfectly absorbing, 150 K grains
around HD 98800 B is
2.4 AU, suggesting a circum-spectroscopic binary distribution.
Our observations set important upper limits on the size of the
inner dust radius of $\sim$2 AU (from the mid-infrared data) and
on the quantity of scattered light of $<$10\% (from the H-band data).  For an
inner radius of 2 AU, the dust distribution must have a height of
at least 1 AU to account for the
fractional dust luminosity of $\sim$20\% L$_B$.
Based on the scattered light limit, the dust grains
responsible for the excess emission must have an albedo of $<$0.33.
The presence of the prominent silicate emission feature at 10$\mu$m
implies dust grain radii of $\ga$2$\mu$m.
The total mass of the dust, located in a circumbinary disk
around the HD 98800 B, is $>$0.002M$_{\earth}$.
The orbital dynamics of the A$-$B pair are
probably responsible for the unusual disk geometry.

%The total mass of the dust is  $>$0.002M$_{\earth}$.
%We conclude that the dust is located in a circumbinary disk
%around the HD 98800 B spectroscopic binary
%with an inner gap of $\sim$2 AU and a height of $\ga$1 AU
%and we speculate that the A$-$B orbital
%dynamics are responsible for the characteristics of
%the observed dust in the system.

\end{abstract}

%% Keywords: journal specific, note !


%% Use the first three characters of the first author's name plus
%% the last two numeral of the year of publication as our KEY for
%% each reference.

\section{Introduction}

HD 98800 is one of the most frequently studied members of
the TW Hydra Association
because of its notably large fractional infrared
(IR) luminosity excess, first detected by
the IRAS satellite \citep{wal88, zuc93}.  
Hipparcos results reported by \citet{fav98} indicate that
HD 98800 is at a distance of 47.6 pc, which has led to
recent age estimates for this K5 system of $\sim$7 Myr,
suggesting that the stars are pre$-$main-sequence (PMS)
(Fekel and Bopp 1993; Soderblom et al. 1996; Webb et al. 1999;
Jensen, Cohen, and Neuh\"auser 1998).
The lack of
obvious signatures of active accretion, such as
hydrogen emission lines \citep{sod96, web99}
and near-IR excess (e.g. Low, Hines, and Schneider 1999), implies that the 
observed mid-IR excess
probably originates in a dusty disk with a central gap.

Complicating both the dynamics and the analysis of the HD 98800
system is its multiplicity.  \citet{inn09} identified it as a visual
binary; its current north-south angular separation of 0.$''$8 corresponds
to a projected linear separation of 38 AU.  Orbital solutions for
the visual binary are not unique because
only linear motion has been observed during the last nine decades.
\citet{tor95} initially found a family of orbits with extremely high
eccentricity ($\sim$0.9), e, and periods, P, $>$10$^5$ years.
Using the Hipparcos data, and with knowledge of the PMS nature
of the system, \citet{tok99} published a different set
of orbital solutions
with moderate 300$-$430 year periods and 0.3$-$0.6
eccentricities.  Tokovinin's orbits
are probably more reliable since they incorporate
better values for the component masses and the
Hipparcos distance, thus improving the determination
of e and P \citep{tor00}.
Spectroscopic measurements 
have shown that each component of the visual binary has a companion.
The southern component, HD 98800 A, is
a single-lined spectroscopic binary (SB)
with P$=$262 days and e$=$0.484,
while the northern component, HD 98800 B,
is a double-lined SB with P$=$315 days and e$=$0.781 \citep{tor95}.
The presence of numerous components raises questions regarding the
relationship between the stars and the dust:
Where does the dust reside?  Is the interaction
between the multiple stellar components and the dust responsible for
HD 98800's unusually large fractional IR excess?  And more
speculatively, can planets form and survive within this system?

To address these questions,
we carried out a 1$-$18$\mu$m diffraction-limited
imaging study
with the W. M. Keck 10-m telescopes, complementing results
presented in several recent papers.
\citet{sod98} and \citet{low99} used, respectively, the WFPC2 and NICMOS
cameras on the {\it Hubble Space Telescope} to
characterize the stellar parameters of
the individual A and B components and to search
for scattered light from the dust.  \citet{geh99} used the ESO
3.6-m telescope to study the
system at 4.7$\mu$m and 9.7$\mu$m,
locating the bulk of the dust around the B component.  \citet{koe00}
combined mid-IR with published millimeter wavelength data,
dramatically illustrating that the excess emission is associated
with HD 98800 B and
modelling the dust distribution.
With a factor of 2$-$4 times higher
spatial resolution than most previous observations, the measurements
in this study are used to characterize the physical properties and spatial
distribution of the dust and stars in HD 98800 and to construct a 
paradigm for the system.
The observations are described in \S2 and the analysis and results in \S3.
A discussion is provided in \S4 and a summary in \S5.

\section{Observations}

\subsection{Mid-Infrared Camera Observations}

HD 98800 (HIP 55505) and the photometric
standard $\mu$ UMa were observed
at the Keck II 10-m telescope on 1998 May 11 (UT)
using OSCIR, the University of Florida
mid-infrared camera.
The camera has
a Rockwell 128$\times$128 Si:As Blocked Impurity
Band (BIB) array with a plate scale of 
0.$''$062/pixel on Keck II, yielding a 7.$''$9$\times$7.$''$9 field of view.
The observations were made through 
light cirrus clouds in four
broad-band filters, K (2.2$\mu$m), M
(3.5$\mu$m), N (10.8$\mu$m) and IHW18 (18.2$\mu$m), and six narrow-band
filters ($\Delta\lambda \sim$1$\mu$m), 7.9$\mu$m, 8.8$\mu$m,
9.8$\mu$m, 10.3$\mu$m, 11.7$\mu$m and 12.5$\mu$m.
An 8$''$ off-source throw
in the north-south direction generated
a standard chop-nod pattern.
Exposures of 30 ms were coadded to produce images at
each chop and nod position.  The chop frequency was 4.1 Hz and
the nod time was approximately 30 seconds.  On-source
integration times were approximately 45 s for all ten filters.
Each nod set was double-differenced to produce background subtracted images.
These differenced images were magnified by a factor of 4, registered
and summed.

\subsection{Near-Infrared Speckle Imaging Observations}

Speckle observations of HD 98800 were carried out at
the Keck I telescope on 1996 January 4$-$6 (UT) using 
the facility 256 x 256 Near-Infrared Camera (NIRC; 
Matthews and Soifer 1994; Matthews et al. 1996), which 
has a plate scale of 0.$''$0203/pixel in its high angular
resolution mode.  Approximately 1,000
short exposures (0.1s) were obtained for HD 98800 and
a nearby SAO star through three filters:
J ($\lambda_o=$1.25$\mu$m),
H ($\lambda_o=$1.65$\mu$m) and K$_{cont}$ ($\lambda_o=$2.2$\mu$m).
Each snapshot was individually sky-subtracted, divided by a flat
field, and corrected for bad pixels,
and then combined with the others at the same wavelength
to construct a diffraction-limited image through standard speckle image 
analysis procedures described in detail elsewhere 
(e.g., Ghez, Neugebauer, and Matthews 1993).

\subsection{Near-Infrared Adaptive Optics Observations}

On 1999 May 26 (UT), HD 98000 and a nearby point source, SAO 180158,
were observed on Keck II using the 
natural guide star adaptive optics
system and KCAM, a 256x256 NICMOS 3 array with a plate scale
of 0.$''$017/pixel \citep{wiz00a, wiz00b}.  HD 98800 proved
to be an excellent adaptive optics guide star, in spite of its
multiple components, and resulted in observations with Strehl ratios of up to
$\sim$0.6 for 5s integrations.  A total of 20 
H-band images were obtained,
interleaved with observations of SAO 180158,
at different positions on the array.  To create
the highest quality final map, each image was magnified by a
factor of two and the 13 frames with
the largest Strehl ratios were combined.

%% In this next section, we use  the \subsection command to set off
%% a subsection.  \footnote is used to insert a footnote to the text.

\section{Data Analysis and Results}

\subsection{The Images}

Figures 1$-$3 show the final diffraction-limited images of HD 98800.  
The resolution ranges from 0.$''$03 at 1.25$\mu$m to 0.$''$46
at 18.2$\mu$m.  The components of the
0.$''$8 north-south pair are easily distinguished out to
12.5$\mu$m.  At 18$\mu$m, the southern component, A, is
marginally detected at $\sim$3$\sigma$.
As originally suggested by \citet{geh99} and confirmed by \citet{koe00}
on the basis of
their mid-IR data, the dust is associated with
the northern component, HD 98800 B.  Between 1$-$18$\mu$m there is no
evidence of dust around HD 98800 A.  Our mid-IR fluxes agree
with those of \citet{koe00} to within 1$\sigma$
at almost every wavelength common to both data sets.

We inspected each component of the wide 0.$''$8 pair for spatially
resolved structure arising from mid-IR thermal dust emission, near-IR
scattered light from the dust, or near-IR photospheric emission from the SB 
components.  Figure 4 is a comparison of the HD 98800 A and B
azimuthal averages at mid-IR wavelengths.  Since HD 98800 A appeared
with HD 98800 B simultaneously in the images at each wavelength, and since no
excess emission was evident in its spectral energy distribution (SED)
(see \S 3.2), HD 98800 A is an ideal point
source for comparison.  At every mid-IR wavelength where the 
signal to noise ratio is 
large enough for a meaningful
comparison of the data, 8.8, 9.8, 10.3 and 11.7$\mu$m,
the HD 98800 B FWHM is comparable to that of HD 98800 A (Figure 4).
This suggests that HD 98800 B is unresolved.
To avoid averaging over structure that is extended
in one particular direction, as might be expected for an inclined
disk, one dimensional profiles
at position angles of 0$^{\circ}$, 45$^{\circ}$, 90$^{\circ}$
and 135$^{\circ}$ were compared; no position angle dependence was detected.
Although the excess emission is unresolved, it is possible
to set upper limits on the
extent of the dust.
A model for HD 98800 B is created from a delta function and a gaussian,
representing, respectively, the stellar
photospheric contribution and the dust component,
where the ratio between the dust and the photosphere
is determined from the SED (see Table 1, \S3.2.1, and Figure 6).
The result is convolved with the PSF (HD 98800 A)
and the azimuthally averaged profile compared to that of HD 98800 B.
We assign a 3$\sigma$ upper limit to the radius of the dust
distribution by identifying the
broadest gaussian that produces a $\Delta\chi^2=$3.
At each of the four mid-IR wavelengths analyzed, this upper limit,
given in Table 1, is close to $\sim$2 AU.

In the J, H, and K-bands,
the spatial resolution is $\sim$4$-$10 times higher than at 10$\mu$m.  
By comparing the azimuthally averaged
AO H-band profiles of HD 98800 A and B and a point source, Figure 5,
it is possible to set a 1$\sigma$ upper limit of $<$10\% on the
reflected light from the dust
around the B component out to a radius of 10 AU.
Given this result and the dust fractional luminosity
of 0.2L$_B$ \citep{zuc93}, we estimate that the grain albedo is less
than 0.33.  There is no evidence
that the HD 98800 A and B subcomponents
are resolved (Figures 2 and 3).  Combining the orbital elements
provided by \citet{tor95} and our estimate of 58$^{\circ}$ for the
inclination of the HD 98800 B system (see \S3.2.2)
to calculate the Ba$-$Bb separation yields $\sim$0.9 AU,
or 0.$''$019, at the time of our J, H, and K speckle observations.

\subsection{Spectral Energy Distributions}

\subsubsection{Construction of SEDs}

Photometry with aperture radii equal to the diffraction
limit provided flux ratios for HD 98800 A and B in the mid-IR and near-IR
AO images.  The flux ratios for the speckle imaging data sets
were obtained from visibility fitting (e.g. Ghez et al. 1993).
Column 7 of Table 2 lists the results.
Absolute photometry from our data sets was compromised by
non-photometric conditions.  Therefore, the unresolved
measurements of \citet{zuc93} provided total flux densities
(column 6 of Table 1).  These values were combined with the flux
ratios to produce the flux densities of the individual components,
which, together with other recent measurements, are listed in
columns 8 and 9 of Table 2.  SEDs for HD 98800 A and B are
plotted in Figure 6.

\subsubsection{Stellar Properties}

The stellar and circumstellar properties
of the HD 98800 system are derived from the
SEDs.
Using the 0.43 $-$ 4.8$\mu$m flux density measurements, we
characterize the photospheric emission with both black body
distributions and Kurucz models obtained with the HST Synphot 
package \citep{kur93}, using log(g)$=$5 and solar metallicity
(see Soderblom et al. 1998).  Table 3 lists the resulting temperatures and
luminosities for both models.  The two approaches yield
comparable quality fits, although, as expected, they differ significantly
in terms of implied temperatures.  The
Kurucz model temperatures agree with those determined
spectroscopically by \citet{sod98} to within 1$\sigma$.  We 
therefore preferentially use the
Kurucz temperatures, 4500 K for
HD 98800 A and 4000 K for HD 98800 B, in our analysis and discussion.

Estimates of the masses and ages for the stellar components are
obtained by comparing their luminosities and temperatures to
PMS evolutionary models.  
For the single-lined spectroscopic
binary HD 98800 A, we consider only Aa,
using our derived Kurucz model temperature and luminosity.
For the double-lined spectroscopic binary HD 98800 B,
the luminosity estimate for HD 98800 A, 0.63L$_{\odot}$,
derived from the Kurucz model (Table 3), is multiplied by
the luminosity ratios given by \citet{tor95}, L$_{Ba}/$L$_A=$0.5
and L$_{Bb}/$L$_A=$0.3, to identify the Ba and Bb component
luminosities.  The individual temperatures for HD 98800 Ba and Bb,
from \citet{sod98}, case B, are 4250 K
and 3700 K, respectively.  With these values for the
temperatures and luminosities, Figure 7 shows the Aa, Ba and Bb
components on the PMS evolutionary tracks of \citet{bar98},
with a mixing length of $\alpha = $1.9, which the recent results of
\citet{whi99} suggest are the most reliable.
These tracks yield
M$_{Aa}=$1.1 $\pm$0.1 M$_{\odot}$, M$_{Ba} = $0.93 $\pm$0.08 M$_{\odot}$
and M$_{Bb} = $0.64 $\pm$0.10 M$_{\odot}$, with an age of
approximately 10$^7$ years for all components (Table 3).

Our mass estimates for HD 98800 Ba and Bb, together with
the mass ratio and the mass functions
from \citet{tor95}, allow us to estimate the inclination
angle, {\it i}, of the spectroscopic binary.
The approach is illustrated in Figure 8,
which shows component mass {\it versus} the inclination angle.
The horizontal bands depict the 1$\sigma$ range for the masses,
obtained from the \citet{bar98} evolutionary tracks, and the curves represent
the dynamically determined M$_1$sin$^3i$ and M$_2$sin$^3i$.  Together these
measurements constrain the HD 98800 Ba$-$Bb inclination angle to be
57.9$^{\circ}$$\pm$0.8.  This procedure was repeated with the tracks of
\citet{swe94}, \citet{dan94} and \citet{dan97} and values
for the inclination of 56.8$^{\circ}$,
63.5$^{\circ}$ and 63.6$^{\circ}$, respectively, were obtained.
The standard deviation of the mean for the four values of inclination
calculated from the different tracks is only 1.8$^{\circ}$.  Therefore, our
estimate of $i_{Ba-Bb}\sim$58$^{\circ}$ is robust.

\subsubsection{Circumstellar Properties}

Figure 9 shows the excess emission, isolated by subtracting
the Kurucz model photosphere from the
HD 98800 B SED.  The
7.9$\mu$m and 12.5$-$100$\mu$m flux densities are used to
fit a black body to the excess, excluding
the prominent silicate emission feature at 10$\mu$m.
A black body of temperature 150$\pm$5 K and
luminosity 0.11$\pm$0.02L$_{\odot}$ gives the
best fit.
Table 4 provides a summary of all the derived dust
properties.  The IRAS and millimeter wavelength data, which do not
resolve HD 98800 A and B, are probably entirely attributable to HD 98800 B,
since in our spatially resolved observations the excess between
7.9 and 18.2$\mu$m is associated solely with the B component.
In our discussion we will assume that this is the case.

\subsubsection{Grain Properties}

As indicated by Skinner, Barlow, and
Justtanont (1992) (see their
Figure 2) and \citet{syl96b}, the emission from the
HD 98800 B dust is consistent with the presence of large grains,
on the order of tens to hundreds of microns in diameter
to account for the observed ratio of infrared to millimeter flux.
With the assumption that the dust is optically thin, we estimate
a lower limit on the size of gravitationally
bound grains stable against blow-out of r$_g>$0.2$\mu$m,
deduced from equating the gravitational and
radiation pressure forces, given a typical ice grain density
of $\sim$1 gm/cm$^3$ \citep{bac93}.
Similarly, for silicate grains, clearly present
around HD 98800 B (Figure 9), $\rho\sim$3 gm/cm$^3$,
and r$_g>$0.07$\mu$m.  However, for silicate grains
to emit efficiently at mid-IR wavelengths, i.e. 2$\pi$r$_g$/$\lambda\la$1,
where $\lambda=$10$\mu$m, they must have r$_g<$1.6$\mu$m
(see http://www.astro.princeton.edu/$\sim$draine/dust/dust.diel.html).  The
Poynting-Robertson lifetime of a $\rho=$3 gm/cm$^3$, r$=$1.6$\mu$m grain
is about 23,000 years.  \citet{zuc93} calculate a
Poynting-Robertson lifetime about five times longer than this
based on typically larger grain sizes.  Assuming that the dust
around HD 98800 B formed at the same time as the stars, which are
estimated to be $\sim$10$^7$ years old, the relatively short
lifetime we calculate
requires some mechanism, such as collisions between
larger bodies, to resupply the small grain population, as has been
pointed out by other authors (e.g., Backman and Paresce 1993).
We estimate a lower bound for the total dust mass
of $\sim$0.002 M$_{\earth}$, given a mass absorption
coefficient of $\sim$1000 cm$^2$/gm and the solid angle for an
annulus of inner radius 2 AU and outer radius 5 AU (see
\S 3.1 and \S 4.1).

\section{Discussion}

\subsection{Disk Structure}

The results presented in \S 3 and in the literature constrain
the structure of the disk around HD 98800 B.
Our estimates for the albedo, $<$0.33, dust temperature, 150 K,
and the total stellar luminosity, 0.58L$_{\sun}$,
allow us to calculate the theoretical equilibrium
radius of the dust.  Because the luminosity
comes from two sources, the stars in the highly eccentric
spectroscopic binary, we make the simplifying assumption here that the
two are equally bright.  A numerical solution for the time 
averaged stellar luminosity as seen from a disk gives a dust
equilibrium radius of $\sim$2.4 AU from the center of mass of the
system.

The criterion of \citet{art94}, [(1/2)$a$(1$+e$)]$+a$,
one half the apastron separation of the stars plus the semi-major
axis, $a$, can be used
to determine the inner radius of a stable, tidally
truncated circumbinary disk.
Combining the data given by \citet{tor95} for $a_1$sin$i$,
$a_2$sin$i$ and the eccentricity, $e$,
with our estimate of $i=$58$^{\circ}$, yields
$a=$1.07 AU; for a stable disk, the inner radius must therefore
be $\ga$2 AU.
The data presented in Table 1 indicate that the 3$\sigma$ upper
observational limit on the inner radius is between 1.8 and
2.1 AU (\S3.1), barely consistent with this result.

From the slight photometric 
variability of the HD 98800 system
seen in the Hipparchos data \citep{sod98},
\citet{tok99} infers that
the Ba$-$Bb binary is seen at an inclination of $\sim$90$^{\circ}$
through the circumbinary dust.
Given the 10$^7$ year age of HD 98800 B,
and the 315 day orbital period of Ba$-$Bb, the dust distribution
and the stellar orbit are probably coplanar (Lubow 2000).
Thus, the 58$^{\circ}$
inclination angle derived for the spectroscopic binary also applies to the
dust.  The large fractional luminosity, 0.2L$_B$,
requires that the dust be vertically extended, since, at a radius of 2 AU,
a thin disk intercepts far too little starlight
to account for the dust brightness.  A disk of perfectly absorbing grains
would have to have a height of at least 1 AU in order to account
for the large fractional luminosity if all the light reradiated by
the disk were intercepted by the grains at a radius of $\sim$2 AU.
Given the upper bound on the albedo, 0.33,
the height of the disk must be between 1 and 1.5 AU.
For a spherical distribution of dust, we expect the fractional
luminosity to be greater than the 0.2L$_B$ observed; thus, a thick disk
geometry is most likely.

The dust around the HD 98800 B spectroscopic binary
is located in a circumbinary disk
with an equilibrium radius of 2.4 AU,
an inner gap of $\sim$2 AU, and a vertical extent
of $>$1 AU.  To account for the observed 8.8$\mu$m
excess flux density of 0.61 Jy, the outer radius of a disk at 47.6 pc with
a 2 AU inner gap and T$=$150 K must be $\sim$5 AU if the origin of the
emission is primarily black body.
(The value listed in Table 2 for the 8.8$\mu$m flux density includes
the stellar photosphere;
Figure 6 shows that the dust to photosphere brightness ratio is 2.1
at 8.8$\mu$m.)
Cooler dust probably extends further out than 5 AU \citep{koe00}, however, it
will eventually be truncated by the dynamical action of the visual
binary's orbit at about one third of the periastron distance between
HD 98800 A and B, i.e. at least $\sim$10$-$15 AU, 
depending on projection
effects.

\subsection{A Paradigm for the HD 98800 System}

We believe that the orbit of the visual binary, HD 98800 A$-$B,
is responsible for the current configuration of the system,
since the inclination of this wide pair is
87$^{\circ}-$89$^{\circ}$
\citep{tor95, tok99}, while the inclination
of HD 98800 Ba$-$Bb and
the associated dust disk is $\sim$58$^{\circ}$.
The A$-$B and the Ba$-$Bb orbits, and therefore the A$-$B orbit and
the B-component dust disk, are not coplanar.
Tidal forces not only may have ripped away
circumbinary material from HD 98800 A in the past,
but also may perturb the HD 98800 B circumbinary disk
in the present, maintaining the small grain population against
Poynting-Robertson drag and giving rise
to its unusual vertical extension.
Disk warping as a result of binary
perturbations from a non-coplanar companion
has been described by \citet{ter96} and \citet{lar96}.  
These perturbations could cause
precession of the HD 98800 B
dust orbital plane, maintaining the 58$^{\circ}$ inclination
constant but
varying the orientation of the disk.  Such changes
would cause the disk to be at times perpendicular
and at times parallel to the A$-$B orbital plane.
We speculate that similar dynamics
destroyed the disk around HD 98800 A,
anomalous in that no evidence of dust is apparent,
and currently maintain the
configuration of the HD 98800 B dust, still prominent
although the stellar ages are about 400 times greater than the
Poynting-Robertson lifetime of the grains, and apparently
significantly vertically extended.
Orbital periods in the models
proposed by \citet{tok99} range from 300 to 430 years,
allowing for relatively short term
periodic perturbations of the HD 98800 B
disk, initiated during periastron passage with HD 98800 A.

%\vfill\eject

\section{Summary}

We have resolved the 0.$''$8 visual binary components in the
young stellar system HD 98800 from 1$-$18$\mu$m,
obtaining complete resolved observations
of the silicate emission feature at 10$\mu$m through
six narrow-band filters, and of the 18.2$\mu$m flux.
The mid-IR excess appears to be azimuthally symmetric
and is entirely associated with
HD 98800 B, a double-lined spectroscopic binary.  The
PSF of HD 98800 B is similar to
that of HD 98800 A in at least the 4 filters spanning 8.8$-$11.7$\mu$m,
and implies an upper limit for the mid-IR emission of $\sim$2 AU.
There is no evidence for any excess emission from HD 98800 B at near-IR
wavelengths, from 1.2$-$4.8$\mu$m.
Our H-band AO images yield an upper bound on the
reflected light from the HD 98800 B excess of $<$10\%
and therefore an albedo for the grains of $<$0.33.

On the PMS tracks of \citet{bar98} the observable
components of HD 98800 appear coeval with an age of
$\sim$10$^7$ years.
Our calculations of the stellar parameters allow us to
estimate new masses for HD 98800 Aa,
Ba and Bb of 1.1, 0.93 and 0.64 M$_{\sun}$, respectively.  From the
masses of the Ba$-$Bb pair we derive $i=$58$^{\circ}$ and, in turn,
a semi-major axis of 1.07 AU.  For the wide visual binary pair,
$i=$90 \citep{tor95, tok99}; 
thus, the orbits of A$-$B and Ba$-$Bb are not coplanar.
A black body fit to the
SED of the HD 98800 B mid-IR excess indicates a dust temperature of 150 K.
If optically thin, the radii of the grains span a range
from $\sim$2$\mu$m through hundreds,
possibly thousands, of microns.
The total mass of the dust is $>$0.002 M$_{\earth}$.  The
Poynting-Robertson lifetime of the smallest grains is $\sim$23,000 
years, suggesting that the disk requires replenishment.

Based on both observational and dynamical considerations,
the inner radius of the dust around HD 98800 B is $\sim$2 AU.  Thus, it must
be a circum-spectroscopic binary structure.  The outer radius
must be at least $\sim$5 AU to account for the observed flux
densities, and cool dust may extend out as far as $\ga$10 AU,
where it is eventually tidally truncated by the orbit of HD 98800 A.
Given the high fractional luminosity, 
the vertical extent of the dust must
be $\sim$1.5 AU.  Thus, the HD 98800 B spectroscopic binary is surrounded
by a thick dust disk, probably in the same orbital plane as
the central binary,
whereas around HD 98800 A, no evidence for any
dust has been detected.  We speculate that, since the orbits of the A$-$B 
pair and the dust disk are not coplanar, dynamical interactions at
the A$-$B orbit periastron are responsible for the survival against
Poynting-Robertson drag and the vertical structure
of the B component disk, as well as for the dissipation of the
A component disk.  Precession of the orientation of the dust plane
resulting from perturbations driven by HD 98800 A may have allowed
various dynamical processes to take place at different times.
It seems unlikely that planetary formation would flourish in such
an environment.  

\vfill\eject

\acknowledgments

We thank the staff and in particular the Observing Assistants, Joel Aycock
and Chuck Sorenson, of the W. M. Keck
Observatory for their logistic and technical support.  We are grateful
to Peter Bodenheimer, Mike Jura, 
Steve Lubow, Mike Simon, Alycia Weinberger,
Mark Wyatt, and Ben Zuckerman
for helpful discussions and suggestions and to
Russel White for producing Figure 7.
We thank Bruce Draine and an anonymous referee
for useful comments on the manuscript.
The data presented here were obtained at the W. M. Keck
Observatory, operated as a scientific partnership between the
California Institute of Technology, the University of 
California, and the National Aeronautics and Space Administration (NASA).
The Observatory was made possible by the contributions of the W. M.
Keck Foundation.  This research was supported by the NASA Origins
of Solar Systems Program grant NAG5-6975 and the Packard Foundation;
additional funding for the AO portion of this work was provided
by the NSF Science and Technology Center for Adaptive Optics.

%%\appendix   %% uncomment if appendices desired...

%%\section{This will be Appendix A, etc.}

%% Reference section....  built in journal names are
%% \apj  \aj  \apjs  \mnras  \pasp  \apjl  \aap   and so on...

%% Only the figure captions, and not the figures
%% themselves, are included in electronic manuscript submissions.

\clearpage

%% No more than seven \figcaption commands are allowed per page,
%% so if you have more than seven captions, insert a \clearpage
%% after every seventh one.

\figcaption{Narrow-band silicate filter images of HD 98800 taken
with the OSCIR mid-IR camera.
North is up and east is to the left.  The
broad-band filter images, K, M, N and IHW18, are not shown.  The
0.$''$8 binary is easily resolved; the airy
ring appears as lobes around the northern component, HD 98800 B.
 \label{fig1}}

\figcaption{From left to right, respectively,
the J, H, and K-band reconstructed speckle images of HD 98800.  North is up
and east is to the left; each frame is $\sim$1.3$''\times$1.3$''$.  
 \label{fig2}}

\figcaption{H-band images of HD 98800 without (left) and
with (right) the Keck adaptive optics (AO) system.  Without AO,
the FWHM of the sources is 0.$''$4.  When AO is used, the FWHM
drops to 0.$''$06.  \label{fig3}}

\figcaption{Azimuthally averaged profiles of HD 98800 A and B at
four narrow-band wavelengths, 8.8, 9.8, 10.3 and 11.7$\mu$m.  The
solid lines join the B component independent
data points, and the dot-dash lines
the A component independent points. \label{fig4}}

\figcaption{Azimuthally averaged AO H-band data for HD 98800 A and B
and the point source SAO 180158.  No significant differences are
apparent in the three profiles.  We use the lower and upper bounds in
the errors for HD 98800 A and B, respectively,
to determine an upper limit on the 
amount of reflected light from the dust in the H-band
around HD 98800 B.  \label{fig5}}

\figcaption{Spectral energy distributions for HD 98800 A, left,
and HD 98800 B, right.  For HD 98800 A,
no excess emission is
evident out to the longest observed 
wavelength of 18.2 $\mu$m; the 3932 K black body is
plotted for reference.  The 3562 K black body fit to HD 98800 B is
also shown.
 \label{fig6}}

\figcaption{The subcomponents HD 98800 Aa, Ba, and Bb plotted
on \citet{bar98} PMS evolutionary tracks.  The 1.0 and 1.5 solar
mass tracks are from models of Palla \& Stahler (1999). \label{fig7}}

\figcaption{Masses of HD 98800 Ba (M$_1$) and Bb (M$_2$) as a function
of orbital inclination of the
Ba$-$Bb spectroscopic binary.  The shaded areas correspond to
1$\sigma$ range for the masses derived from the tracks of \citet{bar98}
and the curved lines represent the 1$\sigma$ range for the mass
functions from \citet{tor95}.  The vertical solid lines
delineate the range of allowed inclination given the
constraints on the masses and mass functions.  \label{fig8}}

\figcaption{The HD 98800 excess emission after subtraction of
the stellar photosphere, represented by Kurucz model photometry.
The data are fit extremely well by a 150 K black body, with
the exception of the silicate emission feature at 
10$\mu$m.  The 0.8, 1.1 and 1.3mm data are from \citet{ruc93},
\citet{syl96}, and Stern, Weintraub, and Festou (1993),
respectively. \label{fig9}}

%\end{document}

%% Tables should be submitted one per page, so put a \clearpage before
%% each one.

\clearpage

\pagestyle{empty}

\begin{deluxetable}{cccc}
\tablewidth{0pt}
\tablecaption{Results of Disk Modelling \label{tbl-4}}
\tablehead{
\colhead{Wavelength ($\mu$m)}   & \colhead{Dust/Photosphere Ratio} & 
\colhead{Radius 3$\sigma$ Upper Limit (AU)}}
\startdata
8.8 & 2.13 & 2.1\\
9.8 & 5.26 & 2.1\\
10.3 & 6.50 & 1.7\\
11.7 & 10.89 & 1.9\\
\enddata

\end{deluxetable}

\clearpage

\landscape
\pagestyle{empty}

\begin{deluxetable}{cccccccccc}
\tablewidth{0pt}
\tablecaption{HD 98800 Flux Densities \label{tbl-1}}
\tablehead{
\colhead{Instrument}     & \colhead{Date}      &
\colhead{Filter}          & \colhead{$\lambda(\mu m)$}  &
\colhead{$\Delta\lambda(\mu m)$}          & \colhead{F$_{\nu}(total)$ (Jy)}  &
\colhead{B/A}  & \colhead{F$_{\nu}$(A) (Jy)}  &
\colhead{F$_{\nu}$(B) (Jy)} &
\colhead{Ref}}
\startdata
WFPC2 &1996 Mar 3 & F439W &0.429 &0.046 &0.391$\pm$0.009 & 0.470$\pm$0.020 & 0.266$\pm$0.008 &0.125$\pm$0.004 & 1\\
WFPC2 &1996 Mar 3 & F555W &0.525 &0.122 & 1.129$\pm$0.025 &0.615$\pm$0.026 & 0.699$\pm$0.021 &0.430$\pm$0.013 & 1\\
WFPC2 &1996 Mar 3 & F953N &0.955 &0.005 & 3.891$\pm$0.083 & 0.945$\pm$0.040 &2.001$\pm$0.060 &1.890$\pm$0.057 & 1\\
NICMOS&1998 May 1 & F095N &0.954 &0.009 & 3.958$\pm$0.084 & 0.904$\pm$0.004 & 2.076$\pm$0.062 & 1.882$\pm$0.056 & 2\\
NICMOS&1997 Jul 3 & F108N &1.082 &0.009 & 4.747$\pm$0.103 & 0.943$\pm$0.007 & 2.443$\pm$0.075 & 2.304$\pm$0.070 & 2\\
NICMOS&'97/'98 & F145M & 1.452 & 0.197 & 4.966$\pm$0.105 & 1.011$\pm$0.002 & 2.470$\pm$0.074 & 2.496$\pm$0.075 & 2\\
NCIMOS&1997 Jul 3 & F187N &1.875 &0.019 & 4.585$\pm$0.099 & 1.091$\pm$0.005 & 2.193$\pm$0.067 & 2.392$\pm$0.073 & 2\\
NICMOS&1998 May 1 & F190N &1.899 &0.017 & 4.705$\pm$0.100 & 1.045$\pm$0.003 & 2.301$\pm$0.069 & 2.404$\pm$0.072 & 2\\
NIRC & 1996 Jan 6 & J & 1.251 & 0.292 & 4.25$\pm$0.09 & 1.096$\pm$0.110 & 2.03$\pm$0.15 & 2.22$\pm$0.15 & 3\\
NIRC & 1996 Jan 5 & H & 1.658 & 0.333 & 4.79$\pm$0.10 & 1.067$\pm$0.107 & 2.32$\pm$0.17 & 2.47$\pm$0.17 & 3\\
NIRC\tablenotemark{a} & 1996 Jan 4 & Kcont & 2.210 & 0.04 & 3.58$\pm$0.11 & 1.002$\pm$0.100 & 1.79$\pm$0.14 & 1.79$\pm$0.14 & 3\\
KCAM$+$AO\tablenotemark{a}&1999 May 25& H& 1.648 & 0.317 & 4.79$\pm$0.10 & 0.991$\pm$0.075 & 2.41$\pm$0.14 & 2.38$\pm$0.14 & 3\\
OSCIR & 1998 May 10 & K & 2.20 & 0.41 & 3.58$\pm$0.11 & 1.001$\pm$0.051 & 1.79$\pm$0.10 & 1.79$\pm$0.10 & 3\\
OSCIR & 1998 May 10 & M & 4.80 & 0.60 & 1.03$\pm$0.05 & 1.282$\pm$0.141 & 0.45$\pm$0.05 & 0.58$\pm$0.06 & 3\\
OSCIR & 1998 May 10&7.9 & 7.90 & 0.76 & 0.61$\pm$0.06 & 2.370$\pm$0.213 & 0.18$\pm$0.03 & 0.43$\pm$0.05 & 3\\
OSCIR & 1998 May 10&8.8 & 8.80 & 0.87 & 1.09$\pm$0.06 & 4.505$\pm$0.183 & 0.20$\pm$0.02 & 0.89$\pm$0.05 & 3\\
OSCIR & 1998 May 10& 9.8& 9.80 & 0.95 & 1.63$\pm$0.08 & 8.929$\pm$0.319 & 0.16$\pm$0.01 & 1.47$\pm$0.08 & 3\\
OSCIR & 1998 May 10&10.3& 10.30& 1.01 & 1.75$\pm$0.09 & 11.905$\pm$0.425 & 0.14$\pm$0.01 & 1.61$\pm$0.09 & 3\\
OSCIR\tablenotemark{a} & 1998 May 10& N  & 10.80& 5.23 & 1.506$\pm$0.151 & 12.821$\pm$0.657 & 0.109$\pm$0.016 & 1.397$\pm$0.145 & 3\\
OSCIR & 1998 May 10&11.7& 11.70& 1.11 & 2.17$\pm$0.11 & 14.286$\pm$0.612 & 0.14$\pm$0.01 & 2.03$\pm$0.11 & 3\\
OSCIR & 1998 May 10&12.5& 12.50& 1.16 & 2.34$\pm$0.12 & 16.129$\pm$1.561 & 0.14$\pm$0.02 & 2.20$\pm$0.12 & 3\\
OSCIR &1998 May 10&IHW18& 18.2 & 1.65 & 5.004$\pm$0.500 & 58.824$\pm$13.841 & 0.084$\pm$0.028 & 4.920$\pm$0.511 & 3\\
TIMMI\tablenotemark{b} & 1996 Apr 5 & M  & 4.71&0.99& 0.90$\pm$0.17   & 1.00$\pm$0.15   & 0.45$\pm$0.09   & 0.45$\pm$0.09 & 4\\
TIMMI\tablenotemark{b} & 1996 Apr 5 & N2 & 9.78&1.29& 1.40$\pm$0.16   & 3.70$\pm$0.56   & 0.30$\pm$0.07   & 1.10$\pm$0.07 & 4\\
IRAS\tablenotemark{a}  & 1983   & 12 & 12 & 7.5 & 2.36$\pm$0.14 & $-$ & $-$ & 2.36$\pm$0.14 &2\\
IRAS  & 1983   & 25 & 25 & 11 &  9.74$\pm$0.78 & $-$ & $-$ & 9.74$\pm$0.78&2 \\
IRAS  & 1983   & 60 & 60 & 40 &  7.05$\pm$0.70 & $-$ & $-$ & 7.05$\pm$0.70&2\\
IRAS  & 1983   & 100& 100& 37 &  4.31$\pm$0.30 & $-$ & $-$ & 4.31$\pm$0.30&2\\
JCMT\tablenotemark{a} & 1992 Feb& 800& 800& $-$&0.111$\pm$0.012 & $-$ & $-$ & 0.111$\pm$0.012&5 \\
JCMT&1993 Apr 25&800& 800& $-$&0.102$\pm$0.010 & $-$ & $-$ & 0.102$\pm$0.010&6\\
JCMT&1992 Feb,Mar&1100&1100&$-$&0.063$\pm$0.006& $-$ & $-$ & 0.063$\pm$0.006&5\\
JCMT\tablenotemark{a}&1994 Apr & 1300&1300& $-$&0.011$\pm$0.003 & $-$ & $-$ & 0.011$\pm$0.003&5\\
IRAM&1993 Apr 17&1300&1300&$-$&0.036$\pm$0.007 & $-$ & $-$ & 0.036$\pm$0.007&7\\
\enddata

%% Any table notes must follow the \end{tabular} command.

\tablenotetext{a}{Data was not used in black body fits or plotted
in figures.}
\tablenotetext{b}{Data plotted in figures but not used in black body fits.}

\tablecomments{Values of F$_{\nu}$(total) for NIRC, KCAM and most OSCIR data 
from ZB93; F$_{\nu}$(total) for OSCIR N and IHW18 filters and all NIRC,
KCAM and OSCIR flux ratios from this work.}

\tablerefs{
(1) Soderblom et al. 1998; K. Noll priv. comm.;
(2) Low et al. 1999; (3) This work; 
(4) Gehrz et al. 1999; (5) Sylvester et al. 1996; (6) Rucinski 1993;
(7) Stern et al. 1993}

\end{deluxetable}

\clearpage

\landscape
\pagestyle{empty}

\begin{deluxetable}{ccccccc}
\tablewidth{0pt}
\tablecaption{HD 98800 Derived Stellar Parameters \label{tbl-2}}
\tablehead{
\colhead{Component}   & \colhead{T$_{BB}$ (K)} & \colhead{T$_{Kurucz}$ (K)} &
\colhead{L$_{BB}$ (L$_{\sun}$)} & \colhead{L$_{Kurucz}$ (L$_{\sun}$)} &
\colhead{R$_{BB}$ (R$_{\sun}$)} & \colhead{M (M$_{\sun}$)}}
\startdata
A (South) &  3932$\pm$5 & 4500$\pm$250 & 0.63$\pm$0.12 & 0.71$\pm$0.14 & 1.71$\pm$0.23 & 1.1$\pm$0.1\tablenotemark{a}\\
B (North) &  3562$\pm$3 & 4000$\pm$250 &0.58$\pm$0.11 &0.58$\pm$0.11 & 2.00$\pm$0.27 & 1.6$\pm$0.1\tablenotemark{b}\\
%Dust &  $-$ & $-$ & $-$ & 2.62$\pm$0.31 AU\tablenotemark{c} \\
%Dust\tablenotemark{a} & 150$\pm$5 & $-$ & 0.11$\pm$0.02& $-$ & 2.19$\pm$0.34 AU\tablenotemark{b}& $-$ \\
%Dust & $-$ & $-$ & $-$ & 2.62$\pm$0.31 AU\tablenotemark{c} \\
\enddata

\tablenotetext{a}{Primary only.}
\tablenotetext{b}{M$_{Ba}\sim$0.93 and M$_{Bb}\sim$0.64 M$_{\sun}$.} 

\end{deluxetable}

\clearpage

\pagestyle{empty}

\begin{table}
\tablewidth{0pt}
\caption{HD 98800 Dust Parameters \label{tbl-3}}
\bigskip
\begin{tabular}{cc}
\tableline\tableline
Blackbody Temperature & 150$\pm$5 K \\
Luminosity & 0.11$\pm$0.02 L$_{\odot}$ \\
Dust Grain Albedo & $<$0.2 \\
Dust Equilibrium Radius & 2.4$\pm$0.3 AU \\
Inner Disk Radius & 1$-$2 AU \\
Outer Disk Radius & 10$-$15 AU \\
\end{tabular}

\end{table}

\clearpage

\begin{figure}
\plotone{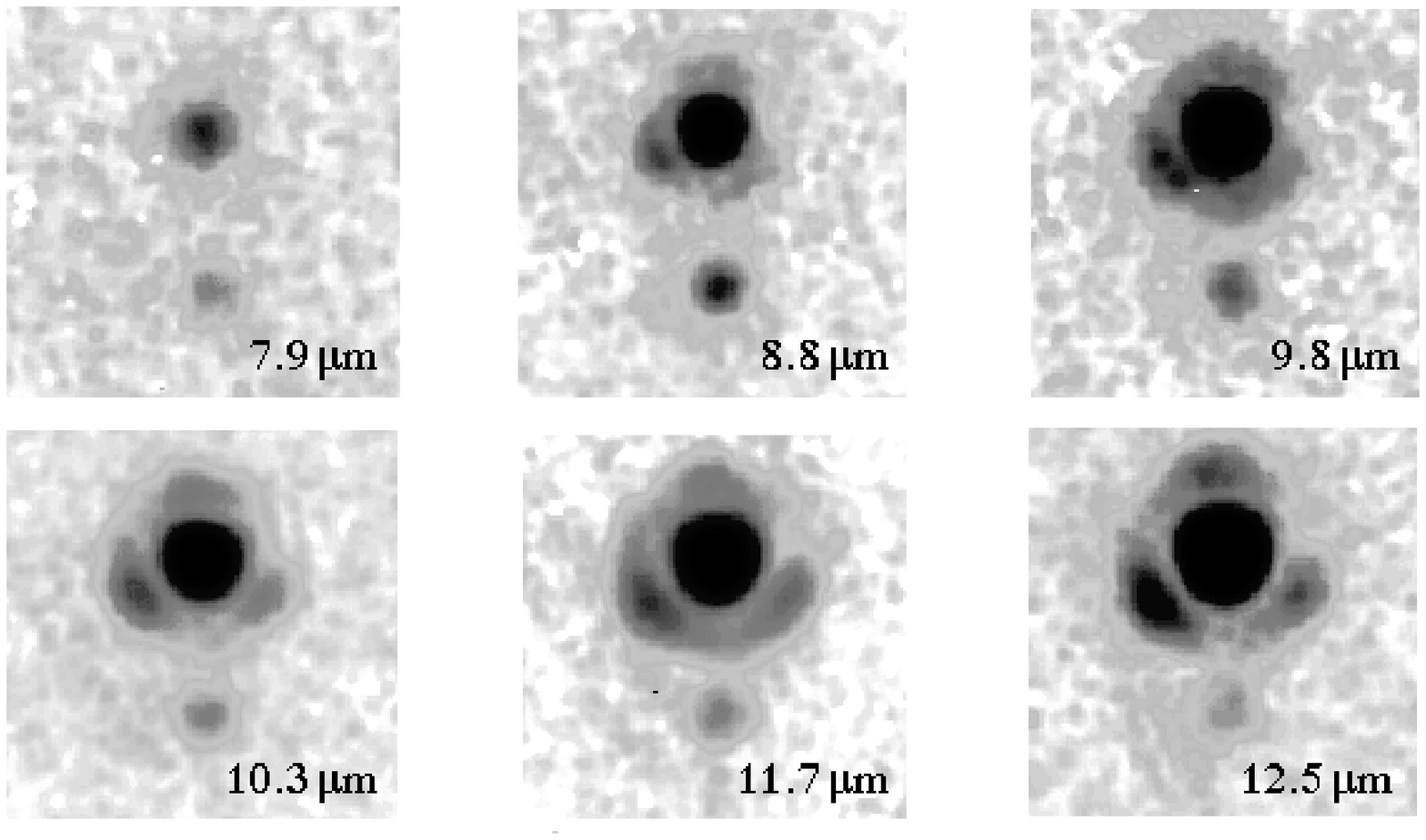}
\end{figure}

\clearpage

\begin{figure}
\plotthree{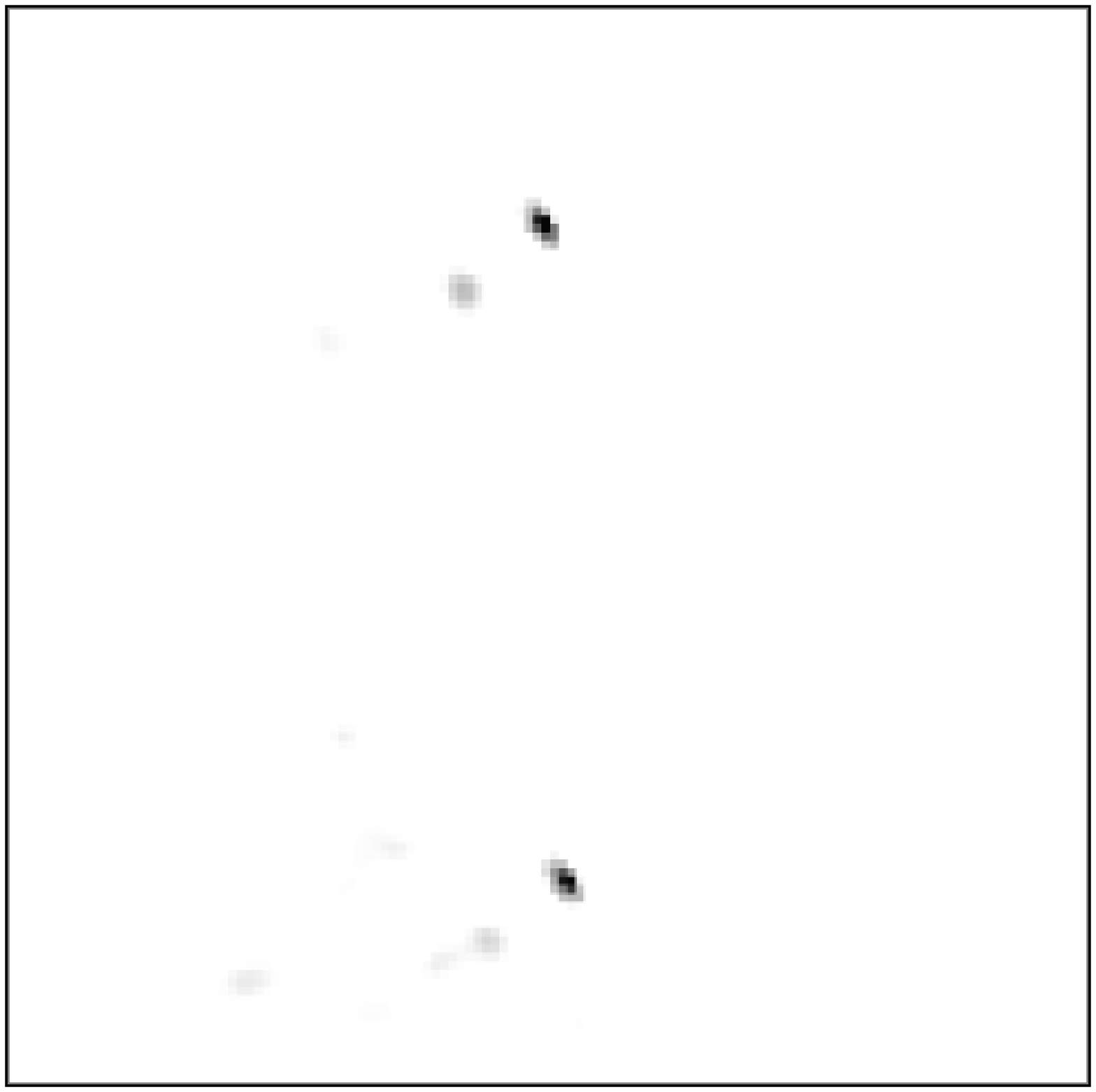}{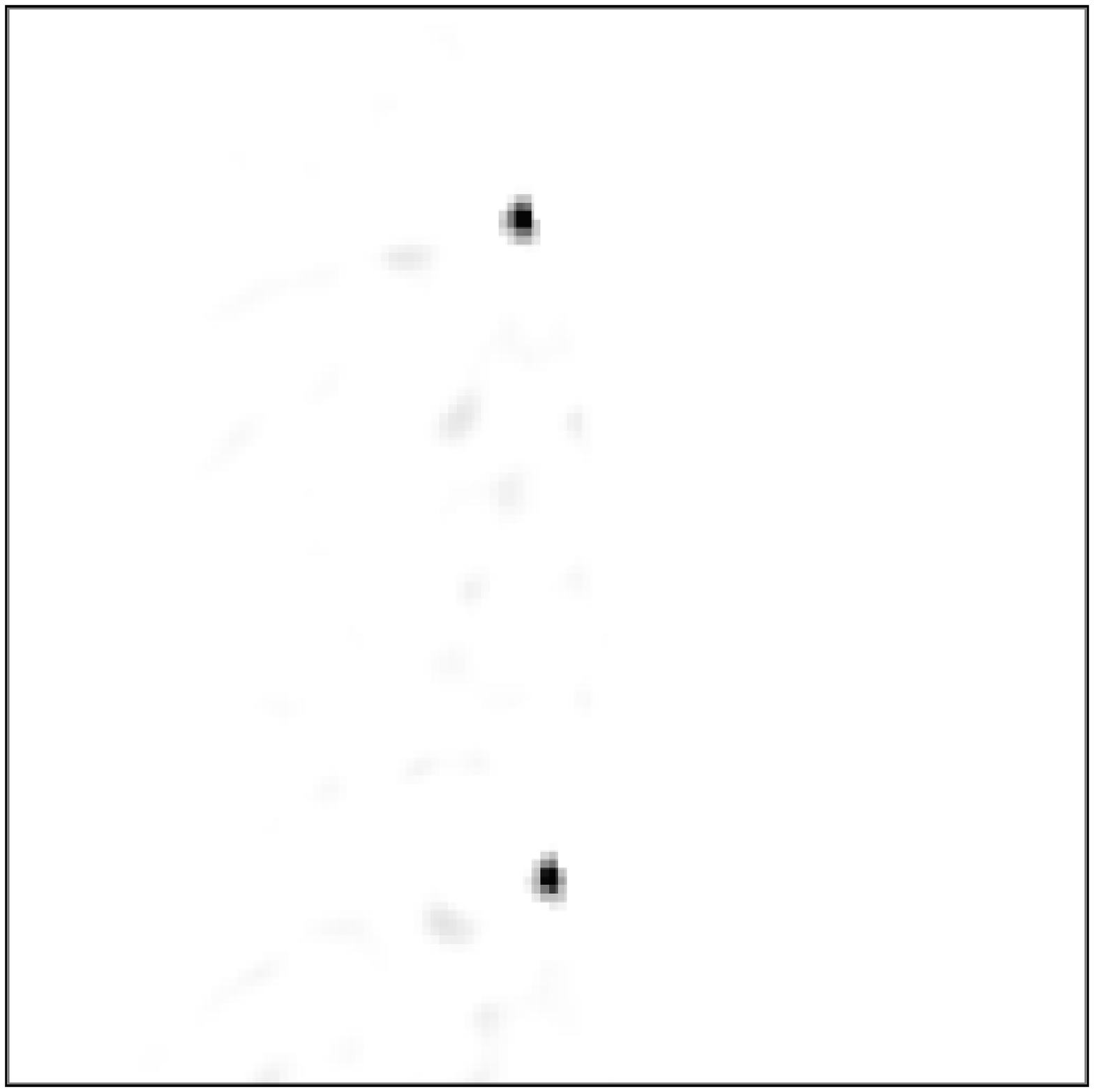}{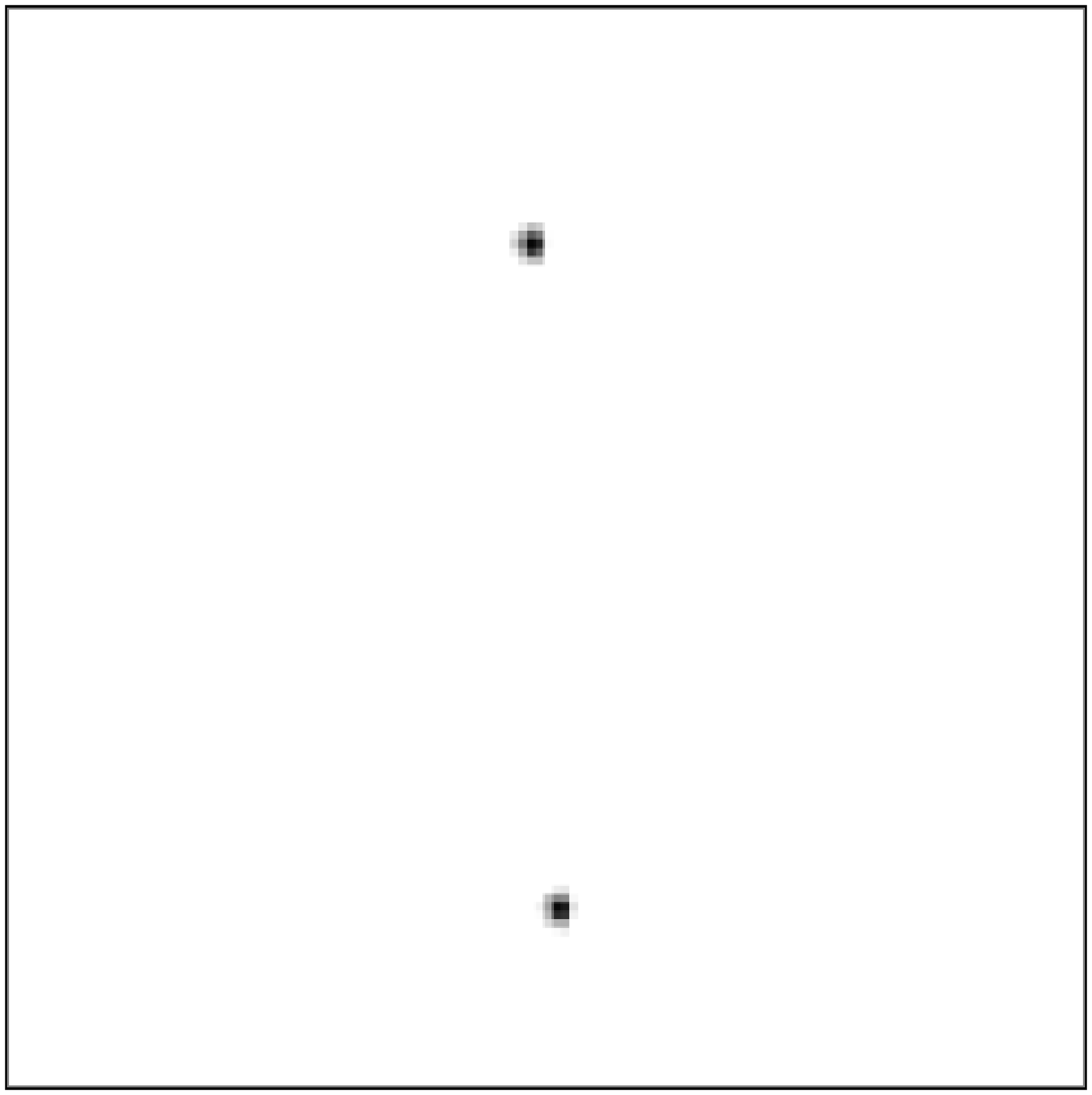}
\end{figure}

\clearpage

\begin{figure}
\plotone{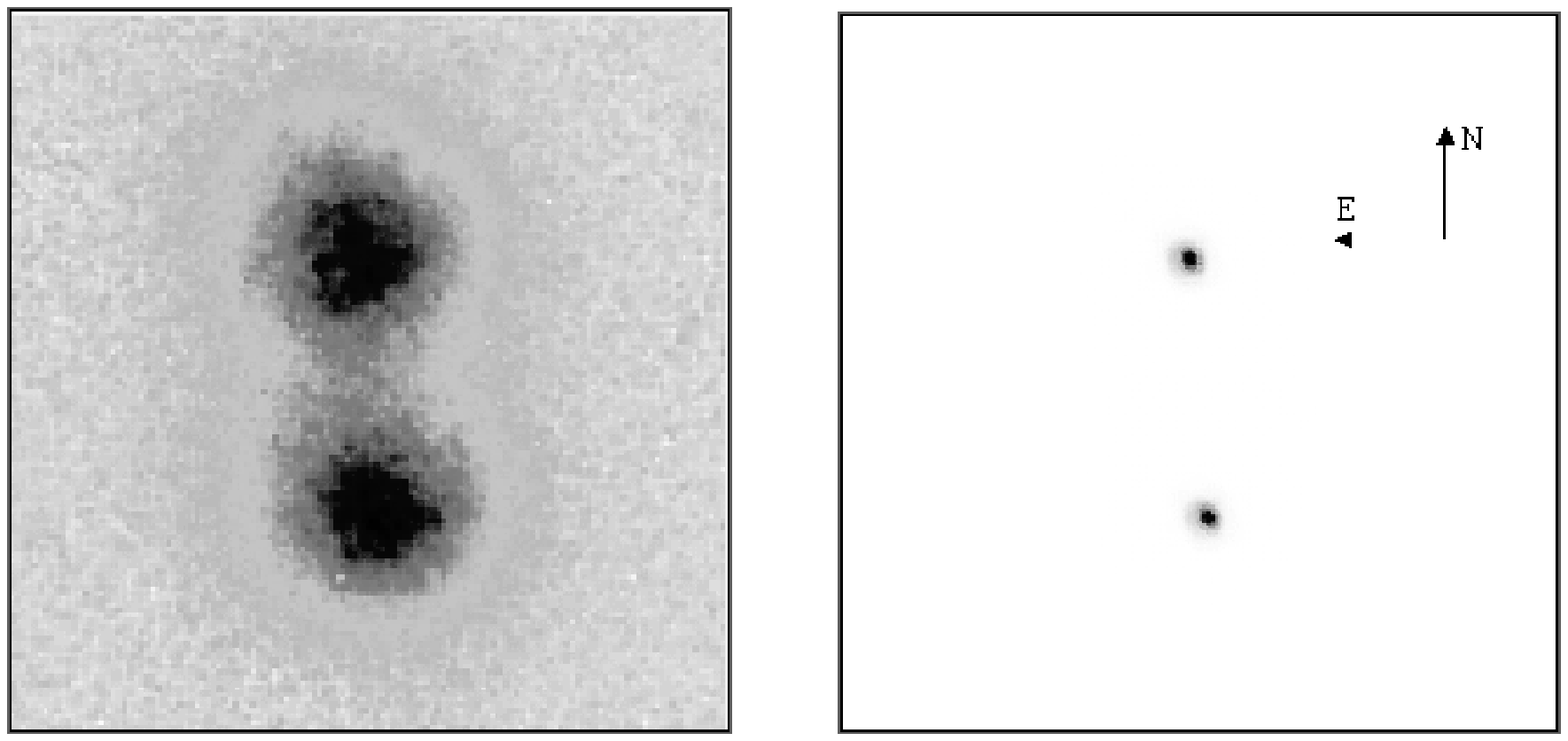}
\end{figure}

\begin{figure}
\plotone{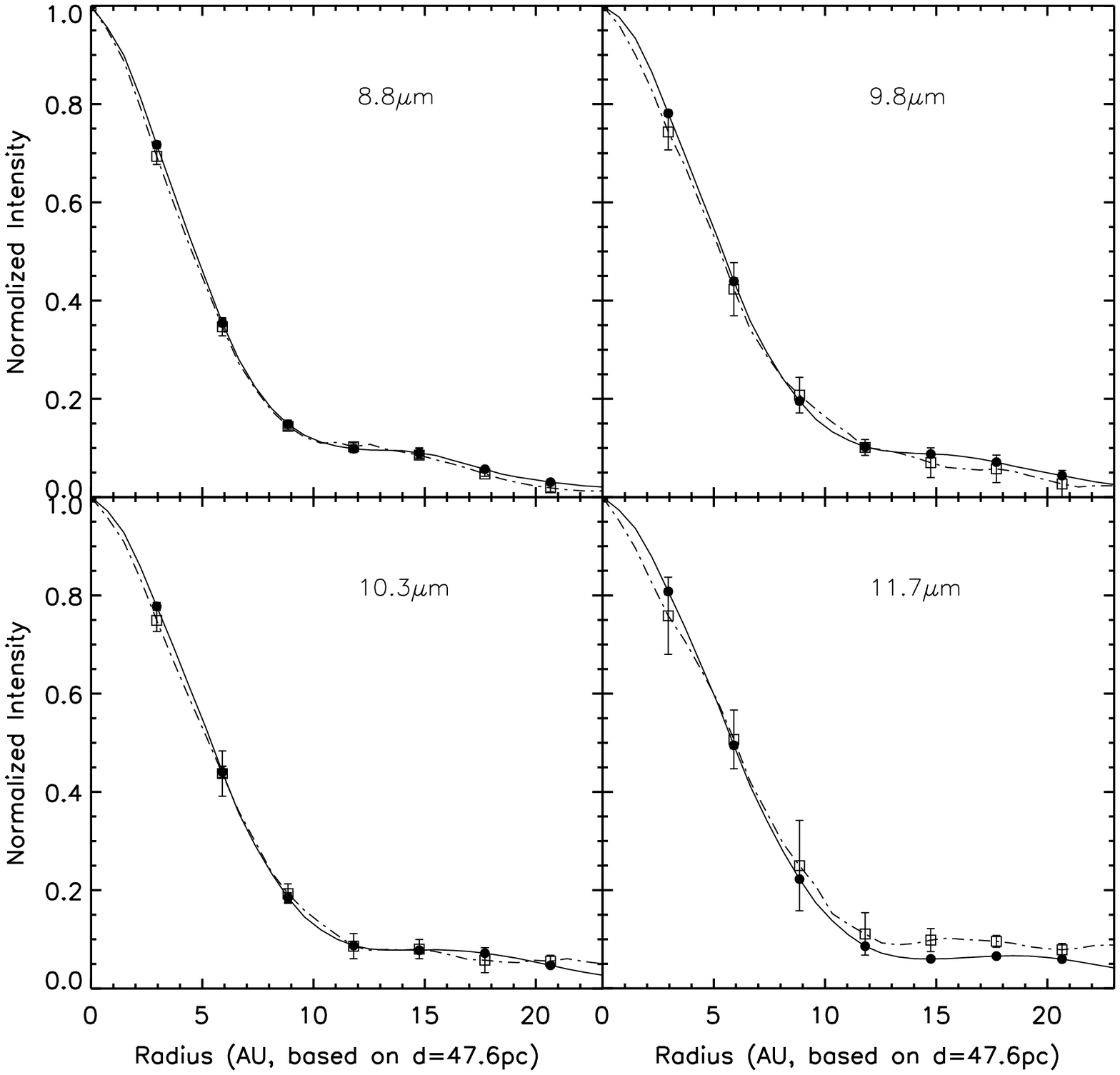}
\end{figure}

\begin{figure}
\plotone{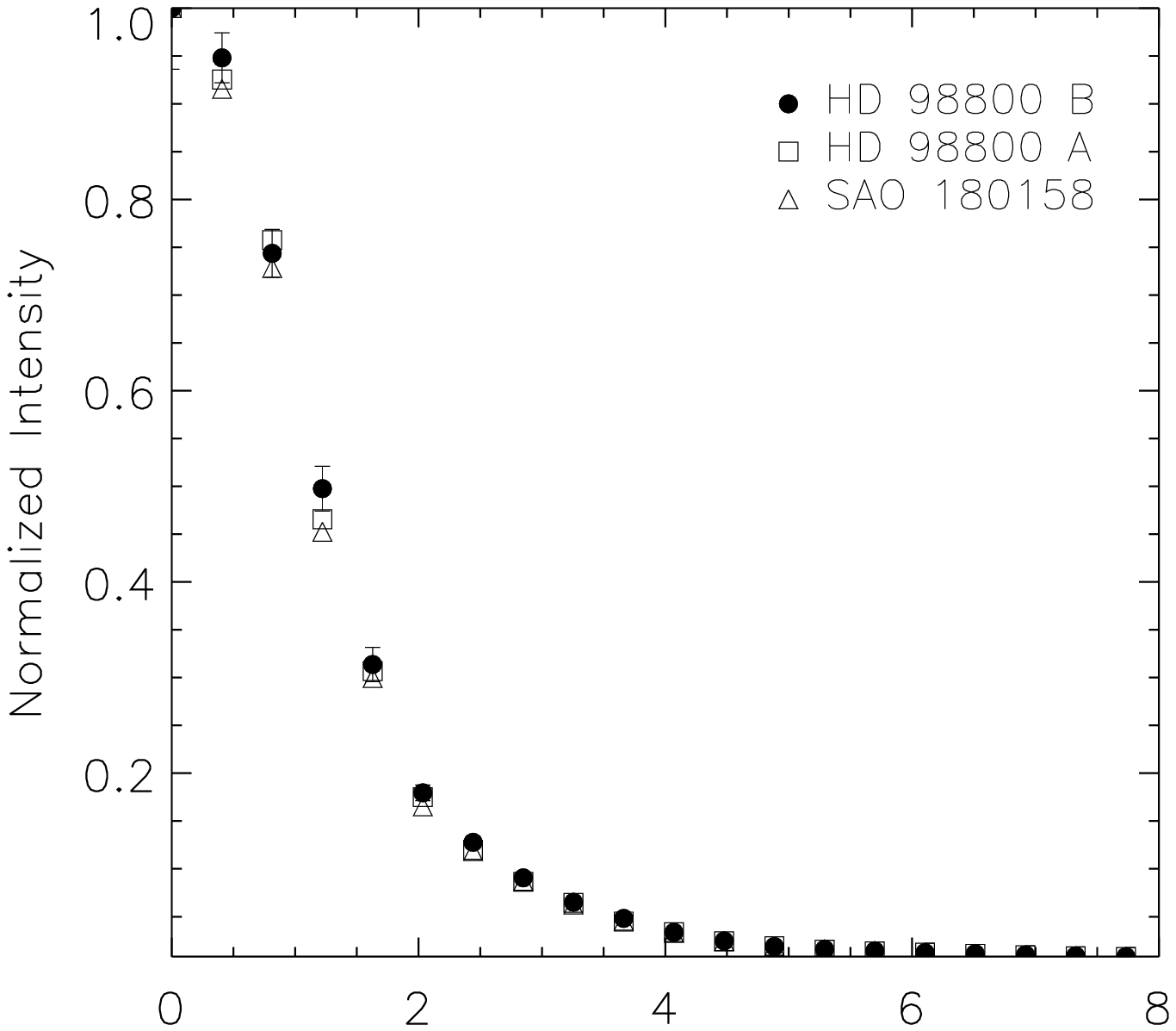}
\end{figure}

\clearpage

\begin{figure}
\plotone{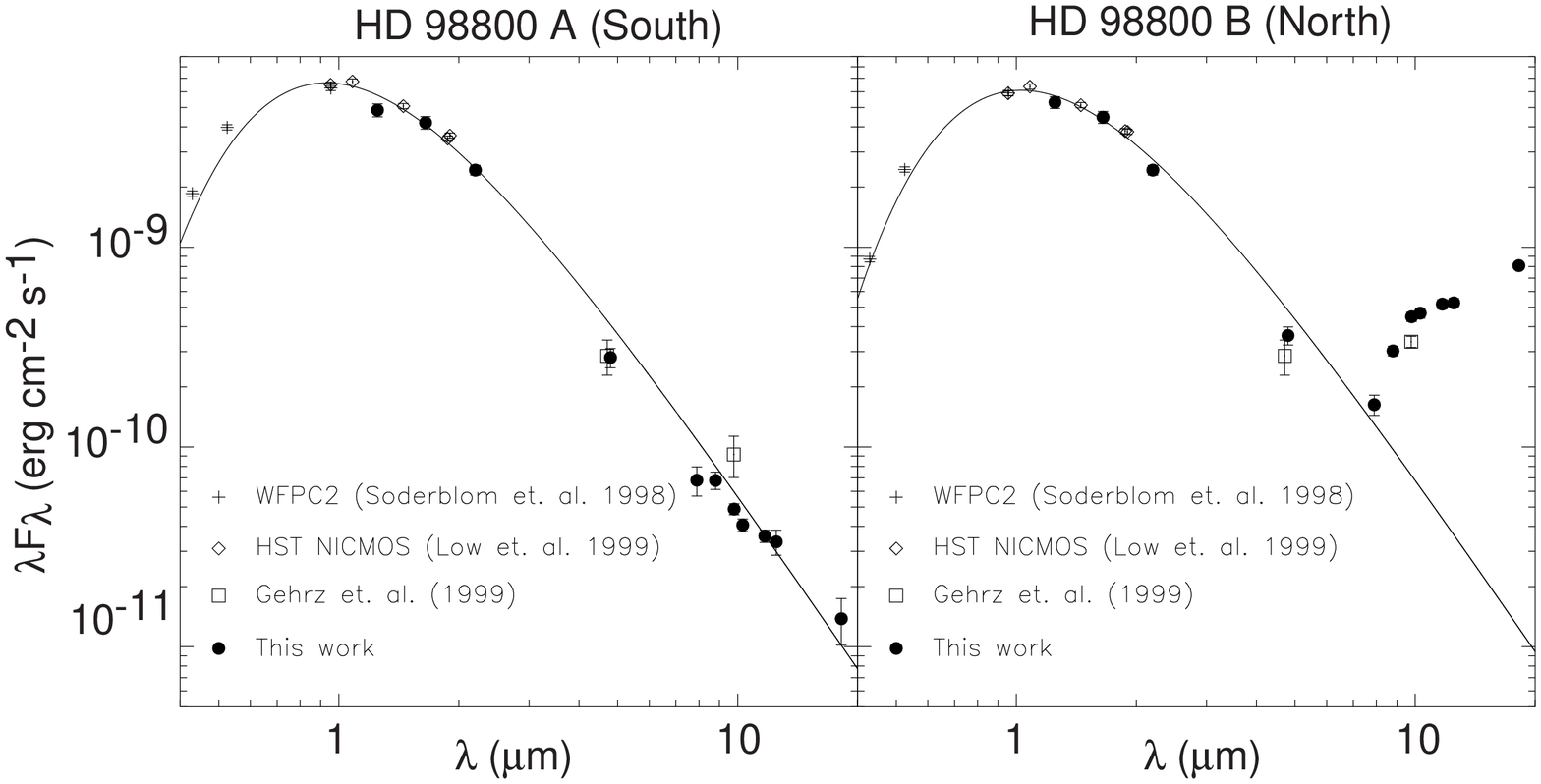}
\end{figure}

\clearpage

\begin{figure}
\plotone{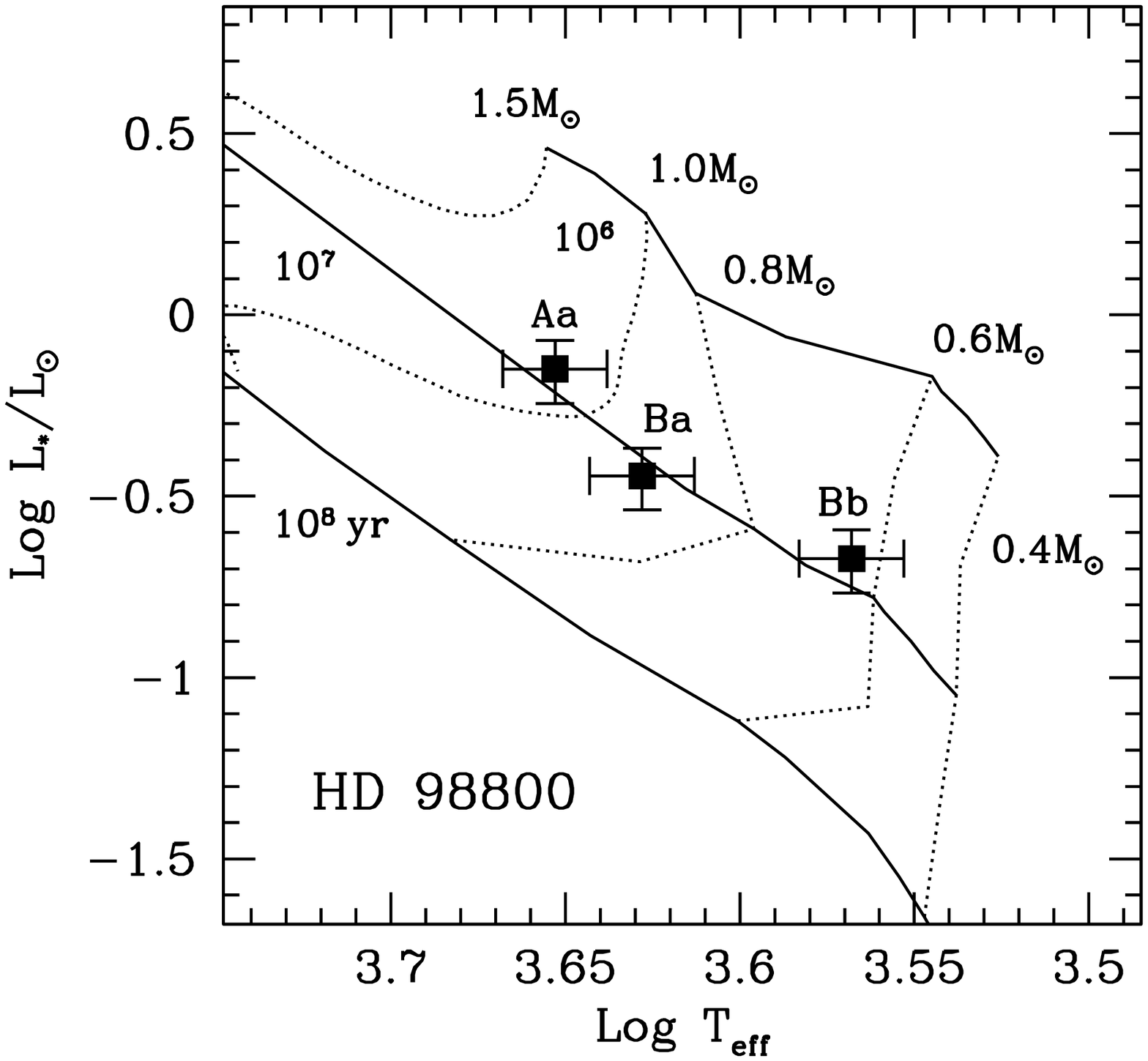}
\end{figure}

\begin{figure}
\plotone{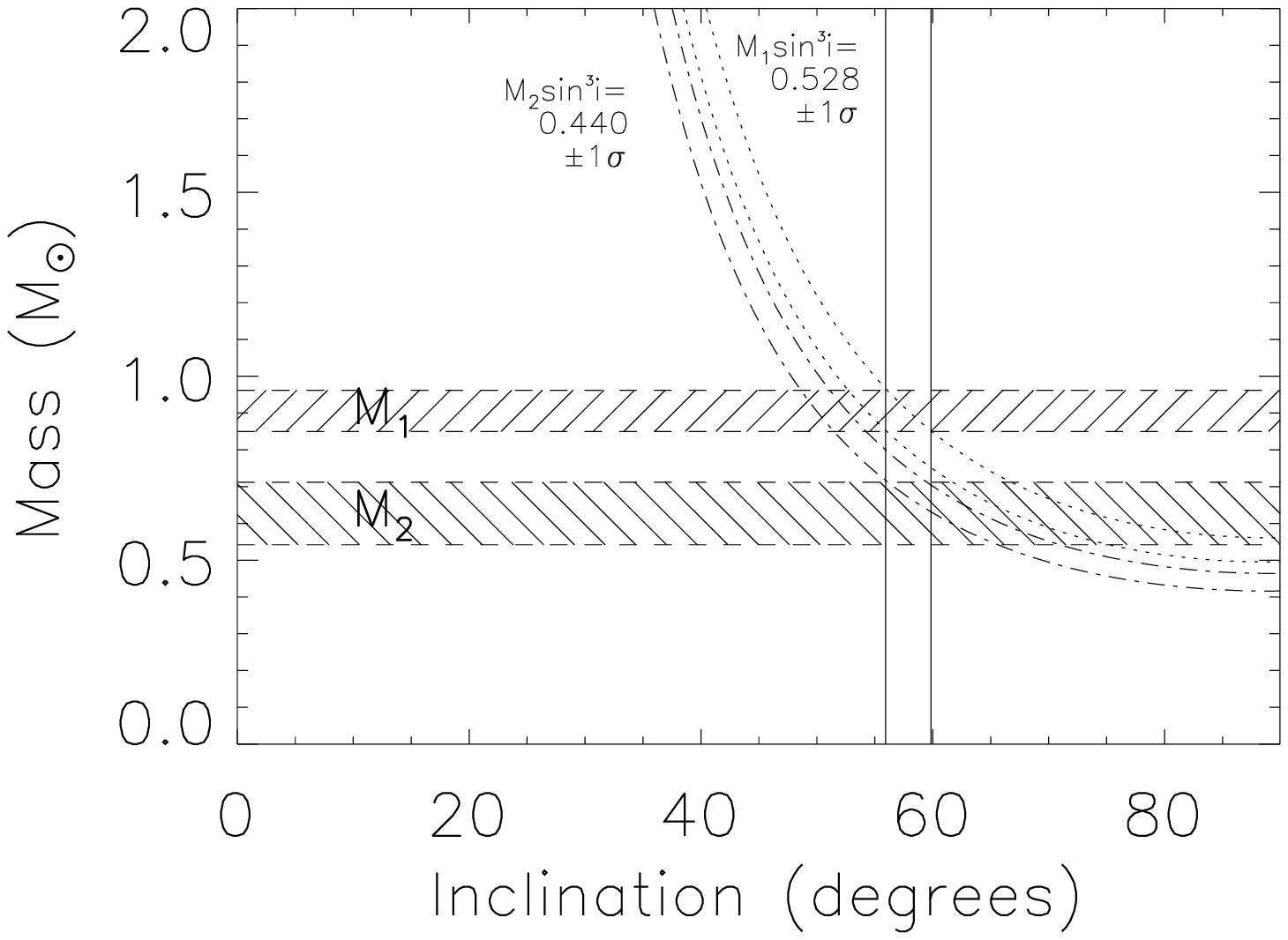}
\end{figure}

\begin{figure}
\plotone{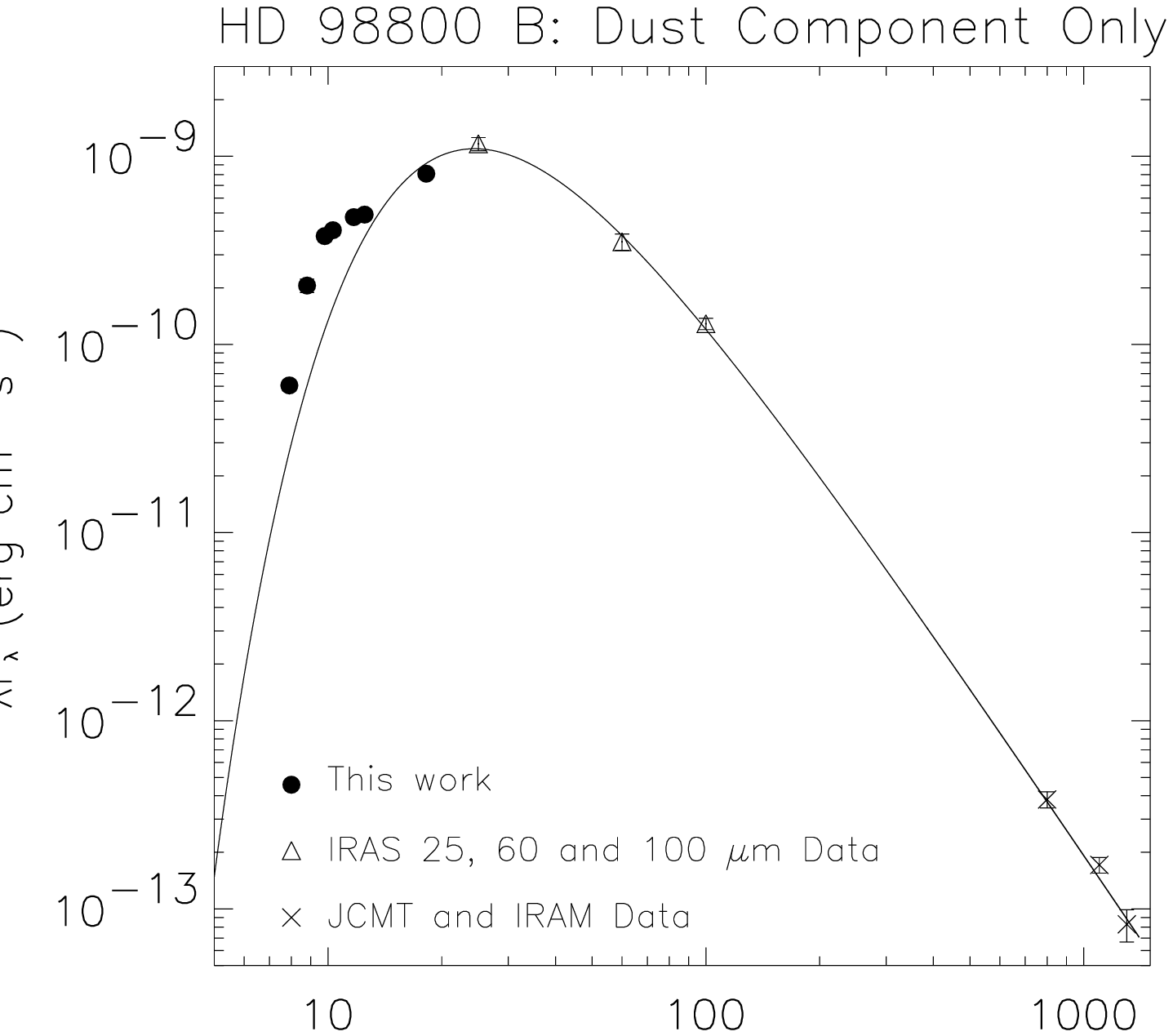}
\end{figure}

\end{document}